\documentclass{ws-procs9x6}
\usepackage{graphicx}

\begin{document}

\title{TOWARDS A GLOBAL ANALYSIS OF \\ GENERALIZED PARTON DISTRIBUTIONS}

\author{K. KUMERI{\v C}KI}

\address{Department of Physics, University of Zagreb, \\
 Zagreb, HR-10002,  Croatia \\
E-mail: kkumer@phy.hr}

\author{D. M\"ULLER\footnote{Talk given by D.M. at the 4th Workshop on {\em Exclusive Reactions at High Momentum Transfer},
May 18-21, 2010,
Thomas Jefferson National Accelerator Facility,
Newport News, Virginia USA.}}

\address{Institut f\"ur Theoretische Physik II, Ruhr-University Bochum,\\
Bochum, D-44780, Germany\\
E-mail: dieter.mueller@tp2.rub.de}

\begin{abstract}
We discuss the complexity of GPD phenomenology, comment on the technological needs for a global analysis,
and report on model and neural network fits to the photon electroproduction off unpolarized proton. We also  point out that Radyushkin's double distribution ansatz is a `holographic' GPD model.
\end{abstract}


\bodymatter

\newcounter{comment}

\newcommand{\comminline}[1]{{%
\refstepcounter{comment}%
\ttfamily\small[$\blacksquare$ \textbf{\underline{Comment}
$\sharp$\thecomment:} #1]}}

\newenvironment{commblock}%
{\refstepcounter{comment}%
\begin{quote}\renewcommand{\baselinestretch}{1}
\ttfamily\small$\blacksquare$ \textbf{\underline{Comment} $\sharp$\thecomment:}}%
{\end{quote}}

\newcommand{\replline}[1]{{
\ttfamily\small[$\blacktriangleright$ \textbf{\underline{Reply}
$\sharp$\thecomment:} #1]}}

\newenvironment{replblock}%
{
\begin{quote}\renewcommand{\baselinestretch}{1}
\ttfamily\small$\blacktriangleright$ \textbf{\underline{Reply} $\sharp$\thecomment:}}%
{\end{quote}}


\section{Introduction}

Generalized parton distributions (GPDs), introduced some time ago,\cite{MueRobGeyDitHor94,Rad96,Ji96a} have received much attention from both the theoretical and the experimental side. This was triggered by the `spin crisis', referring to the mismatch of  quark spin,
extracted from polarized deep inelastic scattering, and the constituent quark model. This is rather analogous to the `momentum crisis',  where almost 50\% of the proton momentum is carried by gluons. Obviously,  both  `crises' of the constituent quark model
are `solved' by switching to the fundamental degrees of freedom.  Hence, we view the `spin puzzle' first as a quest to quantify the partonic structure of the nucleon in terms of quark and gluon angular momenta and second as a wish for an appropriate matching of effective and partonic degrees of freedom. Furthermore, it has been realized that GPDs build up a whole framework for description of hadron structure,\cite{Die03a,BelRad05} including a three dimensional imaging.\cite{RalPir01,Bur00} Certainly,  the  `spin puzzle' is one interesting and intricate aspect, which we will now discuss.

An appropriate decomposition of the nucleon spin in terms of  energy-momentum tensor form factors has been suggested by X.~Ji:\cite{Ji96}
\begin{eqnarray}
\label{JiSR}
\frac{1}{2} =  J^Q + J^G,\quad  J^Q=\sum_{q=u,d,\cdots} J^q , \quad   J^i= A^i(Q^2) + B^i(Q^2),
\end{eqnarray}
where the quark and gluon  contributions are given by  the first moments of parity even  and target helicity (non-)conserved GPD $H$ ($E$),
\begin{eqnarray}
\left\{ A^i \atop B^i\right\}(Q^2) =\lim_{t\to 0,\eta\to 0}\int_{-1}^1\! dx\, \frac{x}{2} \left\{ H^i \atop E^i\right\}(x,\eta,t,Q^2)\,,
\;\; i \in \{u,d,\cdots,G\}\,, \label{eq:AandB}
\end{eqnarray}
taken in the forward limit. Note that the momentum fractions  $A^i$ are obtained from global PDF fits and that  momentum and angular momentum conservation imply that  the total gravitomagnetic moment  vanishes,
$$\sum_{q=u,d,\cdots}  B^q(Q^2)  +B^G(Q^2) =0 \;.$$
Hence, the unknowns in the spin sum rule
are the helicity flip form factors $B^i$. Any `pure' quark model trivially predicts $A^G=0$ and $B^G=0$, whereas taking for  $A^G$  the phenomenological value $\simeq 0.45$ and relying on $B^G=0$, one concludes that quarks and gluons carry roughly the same amount of angular momentum
\footnote{The orbital angular momentum of quarks is defined as $L^q = J^q - \Sigma^q/2$, where $\Sigma^q/2 \sim 0.15$ is the spin of quarks. For more detailed discussions see review.\cite{BurMilNow08}}.
Indeed, lattice calculations indicate a rather small $B^Q$. Unfortunately, these simulations suffer from systematic uncertainties, in particular,  so-called disconnected contributions are presently neglected\cite{Hag09}. Thus,  a phenomenological handle on the spin sum rule is highly desired.

Deeply virtual Compton scattering (DVCS) off nucleon is considered as the theoretically cleanest process offering access to GPDs.
Its amplitude can be parameterized by twelve Compton form factors (CFFs),\cite{BelMueKir01}  which are given in terms of GPDs, e.g., at leading order (LO) in generic notation:
\begin{eqnarray}
\label{DVCS-HS}
\left\{ {\cal H}  \atop {\cal E} \right\}(x_B,t,{\cal Q}^2)  \stackrel{\rm
LO}{=}  \int_{-1}^1\!dx\, \frac{2x}{\xi^2-x^2- i \epsilon}
 \left\{ H \atop E\right\}(x,\eta=\xi,t,{\cal Q}^2)\,.
\end{eqnarray}
The Bjorken variable $x_B$ might be set equal to $2\xi/(1+\xi)$. Favorably, DVCS enters as a subprocess into the hard photon electroproduction where its interference with the Bethe-Heitler bremsstrahlung process provides variety of handles on CFFs. However, the target helicity conserved CFF ${\cal H}$  is often the dominant contribution, while $\cal E$  always appears with a kinematic suppression factor $t/4 M_N^2$, induced by the helicity flip. In other words, one should look at observables for which $ {\cal H}$ is suppressed, too, which requires a transversely polarized target.\cite{Airetal08} Also, in photon electroproduction off neutron\cite{Mazetal07} $ {\cal H}$  is suppressed in the interference term by the accompanying Dirac
form factor $F^n_1$  ($F_1^n(t=0)=0$). Unfortunately, one also has to worry about other non-dominant CFF contributions. Thus, the extraction of  $\cal E$  requires a most complete measurement of all possible observables in dedicated experiments.

Supposing $\cal E$ is measured, the question arises: Can one deconvolute Eq.~(\ref{DVCS-HS})?  Apart from  radiative and higher twist-contributions, one might view the GPD on the $\eta=x$ cross-over line  as a ``spectral function", which provides also the real part of the CFF via a dispersion relation: \cite{Ter05,KumMuePas07,DieIva07,KumMuePas08}
\begin{eqnarray}
\label{DR-Im}
\Im{\rm m}  {\cal F}(x_B,t,{\cal Q}^2)  & \stackrel{\rm LO}{=} & \pi  F (\xi,\xi,t,{\cal Q}^2)\,, \quad F= \{H, E, \widetilde H, \widetilde E\}\,,
\\
\label{DR-Re}
\Re{\rm e}  \!
\left\{\! {\cal H}  \atop {\cal E} \!\right\}\!(x_B,t,{\cal Q}^2) & \stackrel{\rm LO}{=} &
{\rm PV}\! \int_{0}^1\!dx\, \frac{2x}{\xi^2-x^2}\!  \left\{\! H \atop E \! \right\}\! (x,x,t,{\cal Q}^2)
\pm {\cal D}(t,{\cal Q}^2).
\end{eqnarray}
The GPD support properties ensure that Eqs.~(\ref{DR-Im},\ref{DR-Re}) are in {\em one-to-one} correspondence to the perturbative formula (\ref{DVCS-HS}), where  the subtraction constant $\cal D$ can be calculated from either $H$ or $E$. To pin down the GPD in the outer region $y \ge  \eta=x$, one might employ evolution, e.g., in the non-singlet case,
\begin{eqnarray}
\mu^2 \frac{d}{d\mu^2} F(x,x,t,\mu^2) =
 \int_x^1 \frac{dy}{x} V(1,x/y, \alpha_s(\mu)) F(y,x,t,\mu^2)\,.
\end{eqnarray}
A large enough ${\cal Q}^2$ range is not available in fixed target experiments. Hence, we must conclude that essentially only the GPD on the cross-over line (thanks to (\ref{DR-Re}), also outside of the experimentally accessible part of this line,\cite{KumMuePas08}) and the so-called  $D$-term\cite{PolWei99} can be accessed. Moments, such as those entering the spin sum rule (\ref{JiSR}), can only be obtained from a GPD model, fitted to data, or more generally with help of some `holographic' mapping:\cite{KumMuePas08}
\begin{eqnarray}
\label{hol-pro}
\left\{F(x,\eta=0,t,{\cal Q}^2), F(x,x,t,{\cal Q}^2)\right\}  \quad \Longrightarrow \quad   F(x,\eta,t,{\cal Q}^2)  \,.
\end{eqnarray}
Here, $F^i(x,\eta=0,t,{\cal Q}^2)$ are constrained from form factor measurements and, additionally, GPDs $\widetilde H^i$  ($H^i$) by
(un)polarized phenomenological PDFs.

\section{GPD representations and modeling}

Let us now turn to GPD modeling in different representations. First, GPDs might be defined as Radon transform of double distributions:\cite{MueRobGeyDitHor94,Rad97} (DD)
\begin{eqnarray}
F(x,\eta,t,\mu^2) = \int_0^1\! dy\int_{-1+y}^{1-y}\! dz\, (1-x)^p  \delta(x-y-z\eta)  f(y,z,t,\mu^2)\,,
\end{eqnarray}
where $ p \in \{0,1\}.$ In this representation polynomiality, however, not positivity constraints\cite{Pob02} are explicitly implemented. Moreover, with the right choice for $p$,\footnote{Note that the factor $(1-x)$
might be replaced by a more general first order polynomial.} the polynomiality of  $x$-moments can be completed to the required order in $\eta$.
In the central, $-\eta\le x \le \eta$,  and outer, $\eta\le x \le 1$, region the GPD can be interpreted as the probability amplitude of a  $t$-channel meson-like and  $s$-channel parton exchange, respectively. Mathematically, $F$ is a twofold image of the DD $f$, where those in the central and the outer region can be mapped to each other.\cite{MueSch05,KumMuePas07,KumMuePas08} The potential  ambiguity, 
a term that lives only in the central region, is removed by requiring analyticity.\cite{KumMuePas07,KumMuePas08}

Popular GPD models are based on Radyushkin`s DD ansatz\cite{Rad97} for $t=0$,
where the DD factorizes into the PDF analogue $f(y)$ and a normalized profile function $\Pi(z)$.
The GPD on the cross-over line is then given as
\begin{eqnarray}
F(x,x) =  \int_{-1}^1\! \frac{dz}{1-x z} f\left(\frac{x(1-z)}{1-x z}\right) \Pi(z)\,,
\end{eqnarray}
which is a linear integral equation of the first kind within the kernel  $f(\frac{x(1-z)}{1-zx})/(1-x z)$.
Knowing the GPD at $\eta=0$, i.e., $f(y)$, and on the cross-over line,
allows to determine the profile function and so to reconstruct the entire GPD%
\footnote{An example is provided by $f(x) \propto x^{-\alpha} (1-x)^\beta$, which yields after some redefinitions the integral kernel $k = (1-x z)^{-\beta-1}$. The solution is then  obtained in Mellin space.\cite{ManPol98}}, giving example
of the `holographic' mapping (\ref{hol-pro}).

On the first glance a  GPD  in the outer region can be straightforwardly represented by an overlap of light-cone wave functions (LCWF),\cite{DieFelJakKro00BroDieHwa00} which guarantees that positivity constraints are implemented. In simple models one might even reduce the number of non-perturbative functions, e.g.,  in a spectator diquark model, one only deals with one effective scalar LCWF for each struck quark species. This predicts for each of them four chiral even and four chiral odd GPDs. Moreover, one might use such a representation to evaluate also transverse momentum dependent parton distributions (TMDs). Unfortunately, there is a drawback.  In the central region the GPD  possesses an  overlap representation in which the  parton number is not conserved, and where the LCWFs are dynamically tied to those used in the outer region. A closer look reveals that Lorentz covariance already ties the momentum fraction and transverse momentum dependence of a LCWF,\cite{HwaMue07} see also Refs.~\refcite{TibMil01a,MukMusPauRad03,TibDetMil04}. Hence, a overlap representation is only usable if the LCWFs respect Lorentz symmetry, which would allow to restore the GPD in the central region.\cite{HwaMue07}

Strictly spoken, positivity constraints for GPDs are only valid at LO, since they can be violated by the factorization scheme ambiguity. Nevertheless, it would be desired to impose them on GPD models. One might follow the suggestion\cite{Pob02b} and model GPDs  as an integral transform of (triangle) Feynman diagrams, i.e., spectator quark models. A specific integral transformation, namely, a convolution with a spectator mass spectral function,  can be used to include Regge behavior from the $s$-channel view. Such dynamical models provide also effective LCWFs or TMDs; however, simplicity is lost. In particular, PDF and form factor constraints cannot be implemented, i.e., one has to pin down such models within global fitting.

At present we neglect positivity constraints and model GPDs in the most convenient manner by means of a conformal SL(2,$\mathbb{R}$) partial wave expansion, which might be written as a Mellin-Barnes integral\cite{MueSch05}
\begin{eqnarray}
F(x,\eta,t,\mu^2) = \frac{i}{2}  \int_{c -i \infty}^{c+i\infty} \frac{ p_j(x,\eta) }{\sin(\pi j)} F_j(\eta,t,\mu^2)\,.
\end{eqnarray}
Here, $p_j(x,\eta) $ are the partial waves, given in terms of associated Legendre functions of the first and second kind, and the integral conformal GPD moments  $F_j(\eta,t,\mu^2)$
are even polynomials in $\eta$ of order $j$ or $j+1$. The advantages of this representation are: i.~the conformal
moments evolve autonomously at LO, ii.~one can employ conformal symmetry to obtain NNLO corrections to the DVCS amplitude\cite{Mue05aKumMuePasSch06,
KumMuePas07} and iii.~PDF and form factor constraints can be straightforwardly implemented.   Namely,   $F_j(\eta=0,t=0,\mu^2)$ are the Mellin moments of PDFs, $F_{j=0}$ are partonic contributions to elastic form factors,  $H_{j=1}$ and $E_{j=1}$ are the energy-momentum tensor form factors, and for general $j$ one immediately makes contact to lattice measurements. One might expand the conformal moments in terms of  $t$-channel SO(3) partial waves\cite{Pol98} $\hat d_j(\eta)$, expressed by Wigner rotation matrices and normalized to $\hat d_j(\eta=0) =1$. An effective GPD model at given input scale ${\cal Q}^2_0$ is provided by taking into account three partial waves,
\begin{eqnarray}
\label{mod-nnlo}
F_j(\eta,t) =\hat d_j(\eta) f_j^{j+1}(t) +\eta^2 \hat d_{j-2}(\eta) f_j^{j-1}(t) + \eta^4 \hat d_{j-4}(\eta) f_j^{j-3}(t) \,,
\end{eqnarray}
valid for integral $j\ge 4$. Such a model allows us to control the size of the GPD on the cross-over line 
and its ${\cal Q}^2$-evolution, see right panel in Fig.~\ref{fig3}.

\section{Extracting CFFs and GPDs from DVCS measurements}

Certainly, the access to GPDs from experimental data requires some software tools. The variety of both observables and models suggests setting up a flexible architecture, which would allow easy implementation of new models, processes, data sets and fitting strategies. We wrote software prototypes based on object-oriented programming paradigm (using Python) and, alternatively, on functional programming (using Mathematica). In both cases we group scattering processes, theoretical frameworks, models, and experimental data in classes, which also serve as databases. The cross sections are implemented as functions of electromagnetic form factors and CFFs, where models for them are set up separately.  Experimental data are stored in ASCII files, similar to commonly used ones but more standardized. They contain all information needed for the evaluation of observables. After specifying data files, models, and conventions, our prototype software provides theory predictions for data points, depending on model parameters, which can be controlled by a fitting routine.

In a first global fit\cite{KumMue09} to photon electroproduction off unpolarized proton we took sea quark and gluon GPD models with two SO(3) partial waves at small $x$, reparameterized the outcome from H1 and ZEUS DVCS fits at ${\cal Q}^2 = 2\, {\rm GeV}^2$, and employed it in fits of fixed target data within the scaling hypothesis. Thereby, we used the dispersion relation (\ref{DR-Im},\ref{DR-Re}), where
\begin{eqnarray}
\label{ansHval}
H^{\rm val}(x,x,t)  =
\frac{1.35\,  r}{1+x} \left(\frac{2 x}{1+x}\right)^{-\alpha(t)}
\left(\frac{1-x}{1+x}\right)^{b}
\left(1-  \frac{1-x}{1+x} \frac{t}{M^2}\right)^{-1}
\end{eqnarray}
specifies a valence-like GPD on the cross-over line. Here, the skewness
ratio  $r=\lim_{x\to 0}H(x,x)/H(x,0)$, $\alpha(t) =  0.43 + 0.85\, t/{\rm GeV}^2$, $b$  controls the  $x\to1$ limit, and $M$ the residual $t$-dependence, which we set to $M=0.8\,{\rm GeV} $.
The subtraction constant is normalized by $d$ and $M_d$ controls the $t$-dependence:
\begin{eqnarray}
{\cal D}(t) =  d\left(1- \frac{t}{M_d^2}\right)^{-2}\,.
\end{eqnarray}
We also included the parameter-free pion-pole model for the $\tilde E$ GPD  \cite{PenPolGoe99} and parameterized  the $\widetilde H$ GPD rather analogously to Eq.~(\ref{ansHval}) with $b=3/2$.

For the fixed target fits we chose two data sets. The first contains twist-two dominated (preliminary) beam spin asymmetry $A^{(1)}_{\rm BS}$ and beam charge asymmetry  $A^{(i)}_{\rm BC}$ coefficients from HERMES\cite{Ell07,Airetal08} and 12 beam spin asymmetry coefficients $A^{(1)}_{\rm BS}$, which we obtained by Fourier transform of selected CLAS\cite{Giretal07} data with small $-t$.  The second data set includes also  Hall A measurements\cite{Cametal06} for four different $t$ values. In light of the discussion\cite{PolVan08} of Hall A data, we projected on the first harmonic of a {\em normalized} beam spin sum
\begin{eqnarray}
\Sigma_{\rm BS}^{(1),w} =\int_0^{2\pi}\!dw \cos(\phi)  \frac{d\sigma}{dx_{\rm Bj} dt d{\cal Q}^2 d\phi} \Bigg/  \int_0^{2\pi}\!dw  \frac{d\sigma}{dx_{\rm Bj} dt d{\cal Q}^2 d\phi}\,,
\end{eqnarray}
where $dw \propto {\cal P}_1(\phi){\cal P}_2(\phi) d\phi$ includes the Bethe-Heitler propagators, and we also neglected then the helicity dependent cross sections (beam spin differences).
We confirm that  formed beam spin asymmetries are compatible with CLAS ones\cite{Giretal07} and we spell out that the second harmonics in HALL A data, i.e., effective  twist-three contributions, are {\em tiny} or hard to separate from noise. Such contributions are small\footnote{Except for $3\times2$ beam spin asymmetry data points at large $-t$, $x_{\rm B}$, and ${\cal Q}^2$.} in HERMES kinematics, too, where the constant $A^{(0)}_{\rm BC}$, appearing at twist-three level,  is a twist-two dominated quantity that, as expected\cite{BelMueKir01}, turns out to be correlated with  $A^{(1)}_{\rm BC}$.

\begin{figure}[t]
\psfig{file=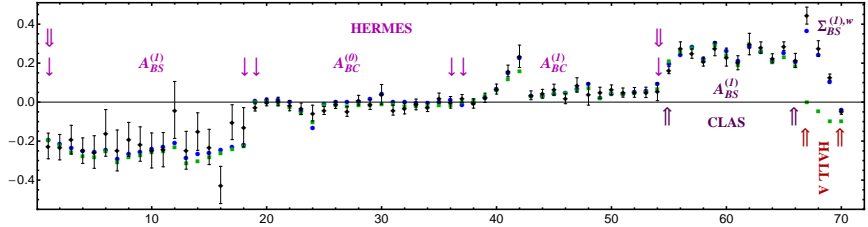, width=11.4cm}%
\vspace{-4pt}
\caption{Experimental vs.~model results\cite{KumMue09}, pinned down by fits without (squares) and with
(circles) HALL A data, for fixed target kinematics: $A^{(1)}_{\rm BS}$ (1-18),  $A^{(0)}_{\rm BC}$  (19-36),
$A^{(1)}_{\rm BC}$ (37-54)
from Ref.~\refcite{Airetal09}; $A^{(1)}_{\rm BS}$  (55-66)
and $\Sigma_{\rm BS}^{(1),w}$  (67-70) are derived from Refs. \refcite{Giretal07} and \refcite{Cametal06}. }
\label{fig1}
\vspace{-4pt}
\end{figure}
To relate the CFFs with the observables we employed the BKM formulas\cite{BelMueKir01} within the `hot-fix' convention\cite{BelMue08}.
First we excluded the Hall A data and set $\widetilde H$ to zero. A least squares fit ($\chi^2/{\rm d.o.f.}\approx 1$)  provides the parameters
\begin{eqnarray}
\label{FitA}
r^{\rm val} =0.95\,,\;\; b^{\rm val}=0.45\,,
\quad
 d=-0.24\,,\;\; M_d=0.5\,{\rm GeV}\,,
\end{eqnarray}
from one local minimum, cf.~description of recent data in Fig.~\ref{fig1} (squares). This should not be considered a unique solution; however, it is compatible with the expectation that the skewness effect at small $x$ should be small, i.e., $r\sim 1$, and that, according to counting rules\cite{Yua03}, $b$ should be smaller than the corresponding $\beta$ value of a PDF. The smallness of $b^{\rm val}$  indicates a rather strong enhancement effect in the resonance region.
In the second fit we included Hall A data, which remains challenging due to the steepness of data points ($-0.33\,{\rm GeV}^2\leq t\leq -0.17\,{\rm GeV}^2$) --- last four points in Fig.~\ref{fig1}.  Again we took some local minimum with $\chi^2/{\rm d.o.f.}\approx 1$, giving
\begin{eqnarray}
\label{FitB}
r^{\rm val} =1.11\,,\;\; b^{\rm val}=2.4\,,
\qquad d=-6.0\,,\;\; M^{\rm sub}=1.5 \,{\rm GeV}\,.
\end{eqnarray}
One observes a slight increase of $r^{\rm val}$  and a larger value of $b^{\rm val}$, i.e., the enhancement of the GPD $H$ in the resonance region diminishes.  In agreement with a chiral quark soliton model estimate,\cite{GoePolVan01} the subtraction constant remains negative and is now sizable; however, at present a positive sign cannot be excluded.   A closer look reveals that our first fit fails to describe Hall A beam spin sums and underestimates the beam spin differences by about 50\%, while our second one still underestimates all the cross sections by about 25\%.
For the latter we find a rather large remainder, effectively parameterized by $\widetilde H$, which is roughly five times bigger than expected. Longitudinally polarized target data provide a handle on $\widetilde H$,\cite{BelMueKir01}  where CFF fits\cite{Gui10} in JLAB kinematics provide at the means a two to three times bigger $\widetilde H$ contribution compared to our expectations ($r_{\widetilde H}\simeq 1, b_{\widetilde H}\simeq2$).
These findings are one to two standard deviations away from our big $\widetilde H$ ad hoc scenario.

\begin{figure}[t]
\psfig{file=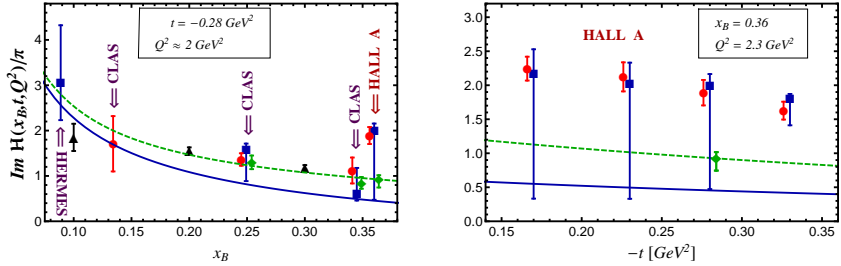, width=11.6cm}
\vspace{-4pt}
\caption{$\Im{\rm m} {\cal H}/\pi$ obtained from different strategies: our model fits\cite{KumMue09} (dashed (solid) curve excludes (includes) Hall A data),
seven-fold CFF fit\cite{Gui08aGuiMou09} with boundary conditions  (squares),  $\cal H$, $\widetilde {\cal H}$ CFF fit\cite{Gui10} (diamonds), smeared conformal partial wave model fit\cite{Mou09} within $H$ GPD (circles). Circles (diamonds) are slightly shifted to the left (right)  hand side.
The triangles result from our neural network fit, cf.~Fig.~\ref{fig3} (left).
}
\vspace{-4pt}
\label{fig2}
\end{figure}
So far we did not study model uncertainties or experimental error propagation, since both tasks might be rather intricate.  To illuminate this, we compare in Fig.~\ref{fig2} our outcomes for $\Im{\rm m}{\cal H}(x_{\rm B},t)/\pi$ versus $x_{\rm B}$ at $t=-0.28\, {\rm GeV}^2$ (left) and for Hall A kinematics $x_{\rm B}$=0.36 versus $-t$ (right)  with results that do provide error estimates. The squares arise from constrained least squares fits\cite{Gui08aGuiMou09} at given kinematic means of HERMES and JLAB  measurements on unpolarized proton, where the imaginary and real parts of twist-two CFFs are taken as parameters. Note that $\Im{\rm m}\widetilde {\cal E}$ and the other remaining eight CFFs are set to zero, however, all available observables, even those which are dominated by these CFFs, have been employed. This might increase `statistics', however, yields also a growth of systematic uncertainties. The huge size of the errors mainly shows to which accuracy $H$ can be extracted from unpolarized proton data alone.\cite{BelMueKir01} A pure $H$ GPD model fit\cite{Mou09} (circles) to JLAB data provides much smaller errors, arising from error propagation and some estimated model uncertainties. Both of our curves are compatible\footnote{Note  that in all fits the unpolarized HALL A cross section at $-t=0.33\, {\rm GeV}^2$ is not well described, see fourth to the last data point in Fig.~\ref{fig1} and the rightmost square in the right panel in Fig.~\ref{fig2}, which results from a  $\chi^2/{\rm d.o.f.}  \approx 2.3$ fit.}  with the findings\cite{Gui08aGuiMou09} and the $H$ GPD model analysis\cite{Mou09} of CLAS data. However, for Hall A kinematics the deviation of the two predictions that are based on $H$ dominance hypothesis, see dashed curve and circles in the right panel, are obvious and are explained by our underestimation of cross section normalization of about 50\%.  Moreover, the quality of fit\cite{Mou09}  $\chi^2/{\rm d.o.f.} \sim 1.7$, might provide another indication that CLAS and Hall A data are not compatible within this hypothesis, see, e.g., the two rightmost circles in the left panel for CLAS
($x_B=0.34$, $t=-0.3 {\rm GeV}^2 $, ${\cal Q}^2=2.3 {\rm GeV}^2$) and Hall A ($x_B=0.36$, $t=-0.28 {\rm GeV}^2 $, ${\cal Q}^2=2.3 {\rm GeV}^2$).
The pure ${\cal H}$ and $\widetilde{\cal H}$ CFF fit\cite{Gui10} (diamonds), including longitudinal polarized target data, is within error bars inconsistent with the $H$ dominated scenario\cite{Mou09} (circles), however, (accidentally) reproduces our dashed curve.   All of these exemplifies that within (strong) assumptions and the present set of measurements the propagated experimental error cannot be taken as an estimate of GPD uncertainties. An error estimation in model fits might be based on twist-two sector projection technique,\cite{BelMueKir01}  boundaries for the superficial model degrees of freedom, and error propagation.

\begin{figure}[t]
\psfig{file=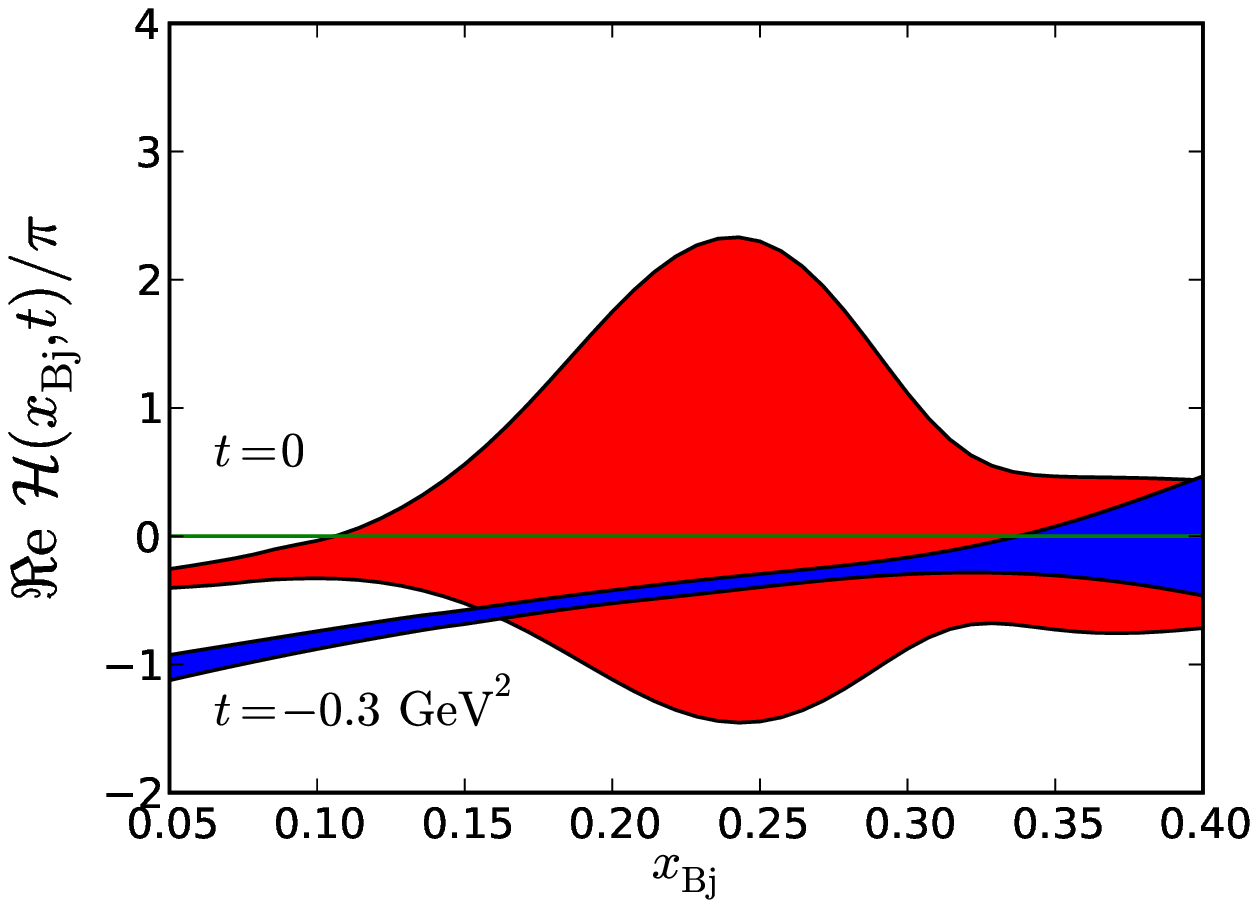, width=5.4cm}%
\hspace*{20pt}\psfig{file=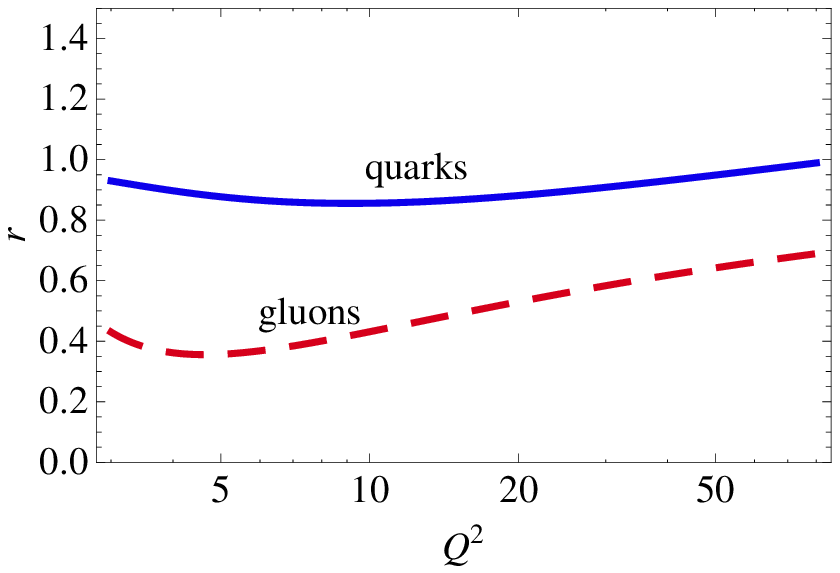, width=5.2cm}%
\vspace{-4pt}
\caption{Neural network extraction of $\Re{\rm e}\,{{\cal H}(x_{\rm Bj},t)}/\pi$ from BCA\cite{Airetal09} and BSA\cite{Giretal07} data (left). Ratio $r=H(x,x)/H(x,0)$ of  LO sea quark and gluon GPDs  for small $x$ versus $Q^2$ (right).   }
\label{fig3}
\vspace{-4pt}
\end{figure}
Let us give a short outlook. Neural networks may be an ideal tool to extract CFFs or GPDs. We present in Fig.~\ref{fig3} a first example in which, within $H$-dominance hypothesis, ${\cal H}$ is extracted by training 50 feed-forward nets with two hidden layers to HERMES BCA\cite{Airetal09} and CLAS BSA\cite{Giretal07} data. Hence, only the experimental errors were propagated, which in absence of a model hypothesis get large for the $t \to 0$  extrapolation. Furthermore, we note that our Fortran code, used so far for small $x$, is now combined  with the dispersion relation model for valence quarks, which gives us the possibility to provide  predictions for the photon electroproduction cross section over a wide kinematical range. Our new fitting results are compatible with the presented ones, where the GPDs are modeled with three effective SO(3) partial waves (\ref{mod-nnlo}).  
Thus, to LO accuracy the skewness ratio for gluons  is now  $r\sim 0.5$ at small $x$  
\footnote{
Note that this rules out the small-$x$ claim,\cite{ShuBieMarRys99} based on incomplete considerations at LO, which would roughly give for quarks and gluons $r\sim 1.6$  and $r\sim 1.1$, respectively. The reader might find further details in Ref.~\refcite{KumMue09b}.}
and  rather stable under evolution, cf.~Fig.~\ref{fig3} and Fig.~4 of Ref.~\refcite{KumMue09}. Finally, we would like to add that the perturbative description of hard exclusive meson electroproduction data is under investigation.\cite{KumLauSch10}.

{\ }

\noindent
In conclusion, a phenomenological  access to the proton spin sum rule can be only reached within an understanding of GPD models, which can be also formulated in terms of an effective nucleon LCWF. A whole framework is available to reveal GPDs and to access the nucleon wave function, which should be considered as the primary task. To do so, reliable data, some mathematical understanding, and appropriate software tools are required. \\{\ }

\noindent
{\bf Acknowledgments}\\

\noindent
D.M. is indebted to P.~Stoler and A.~Radyushkin for invitation to the workshop {\em Exclusive Reactions at High Momentum Transfer}.
We are grateful to T.~Lautenschlager, K.~Passek-Kumeri{\v c}ki, A.~Sch\"afer, and Z.~Vlah for many fruitful discussions. We also like to thank H.~Moutarde for general discussions on software architecture.  Both K.K.~and D.M.~like to thank the Theory Group at the University of Regensburg for the warm hospitality during final stages of the work. This work was supported by the Croatian Ministry of Science, Education and Sport, contract no. 119-0982930-1016, and by the German Research Foundation contract DFG 436 KRO 113/11/0-1.

\end{document}